\documentstyle[12pt]{article}

\newcommand{\sect}[1]{\setcounter{equation}{0}\section{#1}}

\def\bseq{\begin{subequation}}  
\def\eseq{\end{subequation}}
\def\bsea{\begin{subeqnarray}}  
\def\esea{\end{subeqnarray}}

\newcommand{\beq}{\begin{equation}}
\newcommand{\eeq}{\end{equation}}
\newcommand{\bea}{\begin{eqnarray}}
\newcommand{\eea}{\end{eqnarray}}
\newcommand{\non}{\nonumber}
\def\bq{\begin{quote}}
\def\eq{\end{quote}}

\newcommand{\EQ}{\begin{equation}}
\newcommand{\EN}{\end{equation}}
\newcommand{\ena}{\end{eqnarray}}

\renewcommand{\a}{\alpha}
\renewcommand{\b}{\beta}

\renewcommand{\d}{\delta}
\newcommand{\th}{\theta}

\newcommand{\pa}{\partial}
\newcommand{\di}{\partial}
\newcommand{\g}{\gamma}
\newcommand{\G}{\Gamma}

\renewcommand{\l}{\lambda}
\renewcommand{\L}{\Lambda}

\newcommand{\f}{\phi}

\newcommand{\x}{\chi}

\newcommand{\s}{\sigma}
\renewcommand{\S}{\Sigma}

\renewcommand{\O}{\Omega}

\newcommand{\Db}{\bar{D}}

\newcommand{\Fb}{\bar{F}}

\newcommand{\phib}{\bar{\phi}}
\newcommand{\fb}{\bar{\phi}}
\newcommand{\gb}{\bar{\g}}
\newcommand{\Lb}{\bar{\Lambda}}
\newcommand{\Lmb}{\bar{\Lambda}}

\newcommand{\calH}{{\cal H}}
\newcommand{\calHt}{\tilde{{\cal H}}}
\newcommand{\calHpp}{{{\cal H}}^\pp}
\newcommand{\calHtpp}{{\tilde{{\calH}}^\pp}}

\newcommand{\xpp}{\chi^\pp}
\newcommand{\xb}{\bar{\chi}}
\newcommand{\xbpp}{{\xb}^\pp}

\def\Mb{\kern 2pt\mathchoice
            {
             \vbox{\hrule width10pt height 0.4pt depth 0pt
                 \kern 1.2pt\hbox{\kern -2pt$\displaystyle M$}}}
            {
                 \vbox{\hrule width10pt height 0.4pt depth 0pt
                 \kern 1.2pt\hbox{\kern -2pt$\textstyle M$}}}
            {
\vbox{\hrule width6pt height 0.4pt depth 0pt
                 \kern 1.0pt\hbox{\kern -2pt$\scriptstyle M$}}}
            {
                 \vbox{\hrule width5pt height 0.4pt depth 0pt
                 \kern 0.8pt\hbox{\kern -2pt$\scriptscriptstyle M$}}}}

\def\Sb{\kern 2pt\mathchoice
            {
                 \vbox{\hrule width6pt height 0.4pt depth 0pt
                 \kern 1.2pt\hbox{\kern -2pt$\displaystyle S$}}}
            {
                 \vbox{\hrule width6pt height 0.4pt depth 0pt
                 \kern 1.2pt\hbox{\kern -2pt$\textstyle S$}}}
            {
                 \vbox{\hrule width3.5pt height 0.4pt depth 0pt
                 \kern 1.0pt\hbox{\kern -2pt$\scriptstyle S$}}}
            {
                 \vbox{\hrule width3pt height 0.4pt depth 0pt
                 \kern 0.8pt\hbox{\kern -2pt$\scriptscriptstyle S$}}}}

\def\Rb{\kern 2pt\mathchoice
            {
                 \vbox{\hrule width5.5pt height 0.4pt depth 0pt
                 \kern 1.2pt\hbox{\kern -2.5pt$\displaystyle R$}}}
            {
                 \vbox{\hrule width5.5pt height 0.4pt depth 0pt
                 \kern 1.2pt\hbox{\kern -2.5pt$\textstyle R$}}}
            {
                 \vbox{\hrule width3.5pt height 0.4pt depth 0pt
                 \kern 1.0pt\hbox{\kern -2.2pt$\scriptstyle R$}}}
            {
                 \vbox{\hrule width3pt height 0.4pt depth 0pt
                 \kern 0.8pt\hbox{\kern -2.2pt$\scriptscriptstyle R$}}}}

  \def\pp{{\mathchoice
          {
              \kern 1pt%
              \raise 1pt
              \vbox{\hrule width5pt height0.4pt depth0pt
                    \kern -2pt
                    \hbox{\kern 2.3pt
                          \vrule width0.4pt height6pt depth0pt
                          }
                    \kern -2pt
                    \hrule width5pt height0.4pt depth0pt}%
                    \kern 1pt
           }
            {
              \kern 1pt%
              \raise 1pt
              \vbox{\hrule width4.3pt height0.4pt depth0pt
                    \kern -1.8pt
                    \hbox{\kern 1.95pt
                          \vrule width0.4pt height5.4pt depth0pt
                          }
                    \kern -1.8pt
                    \hrule width4.3pt height0.4pt depth0pt}%
                    \kern 1pt
            }
            {
              \kern 0.5pt%
              \raise 1pt
              \vbox{\hrule width4.0pt height0.3pt depth0pt
                    \kern -1.9pt  
                    \hbox{\kern 1.85pt
                          \vrule width0.3pt height5.7pt depth0pt
                          }
                    \kern -1.9pt
                    \hrule width4.0pt height0.3pt depth0pt}%
                    \kern 0.5pt
            }
            {
              \kern 0.5pt%
              \raise 1pt
              \vbox{\hrule width3.6pt height0.3pt depth0pt
                    \kern -1.5pt
                    \hbox{\kern 1.65pt
                          \vrule width0.3pt height4.5pt depth0pt
                          }
                    \kern -1.5pt
                    \hrule width3.6pt height0.3pt depth0pt}%
                    \kern 0.5pt
            }
        }}

  \def\mm{{\mathchoice
                       {
                             \kern 1pt
               \raise 1pt    \vbox{\hrule width5pt height0.4pt depth0pt
                                  \kern 2pt
                                  \hrule width5pt height0.4pt depth0pt}
                             \kern 1pt}
                       {
                            \kern 1pt
               \raise 1pt \vbox{\hrule width4.3pt height0.4pt depth0pt
                                  \kern 1.8pt
                                  \hrule width4.3pt height0.4pt depth0pt}
                             \kern 1pt}
                       {
                            \kern 0.5pt
               \raise 1pt
                            \vbox{\hrule width4.0pt height0.3pt depth0pt
                                  \kern 1.9pt
                                  \hrule width4.0pt height0.3pt depth0pt}
                            \kern 1pt}
                       {
                           \kern 0.5pt
             \raise 1pt  \vbox{\hrule width3.6pt height0.3pt depth0pt
                                  \kern 1.5pt
                                  \hrule width3.6pt height0.3pt depth0pt}
                           \kern 0.5pt}
                       }}

\def\pd{{\kern0.5pt
                   + \kern-5.05pt \raise5.8pt\hbox{$\textstyle.$}\kern
0.5pt}}

\def\pmd{{\kern0.5pt
                  \pm \kern-5.05pt \raise6.3pt\hbox{$\textstyle.$}\kern1.5pt}}

\def\md{{\mathchoice
   {
      {{\kern 1pt - \kern-6.2pt \raise5pt\hbox{$\textstyle.$}\kern 1pt}}}
    {
      {{\kern 1pt - \kern-6.2pt \raise5pt\hbox{$\textstyle.$}\kern 1pt}}}
    {
      {\kern0.5pt - \kern-5.05pt \raise3.4pt\hbox{$\textstyle.$}\kern0.5pt}}
    {
      {\kern0.5pt - \kern-5.05pt \raise3.4pt\hbox{$\textstyle.$}\kern0.5pt}}}}

\newcommand{\Eh}{\hat{E}}

\newcommand{\Apm}{A_+^{~-}}
\newcommand{\Amp}{A_-^{~+}}

\newcommand{\BAH}{\buildrel \leftarrow \over H}

\newcommand{\dirl}{\buildrel \leftrightarrow \over \di_{\pp}}

\newcommand{\ad}{{\dot{\alpha}}}

\newcommand{\Del}{\nabla}
\newcommand{\Delb}{\bar{\nabla}}

\newcommand{\Delm}{\nabla_{-}}
\newcommand{\Delpd}{\nabla_{\pd}}

\def\Sc{\scriptstyle}

\newcommand{\reff}[1]{(\ref{#1})}

\newcommand{\sihalf}{{\Sc\frac{i}{2}}}

\newcommand{\half}{\frac{1}{2}}
\newcommand{\ihalf}{\frac{i}{2}}

\renewcommand{\thefootnote}{\fnsymbol{footnote}}

\begin{document}

\newpage
\begin{titlepage}
\begin{flushright}
{hep-th/9509103}\\
{BRX-TH-382}
\end{flushright}
\vspace{2cm}
\begin{center}
{\bf {\large QUANTUM (2,2) SUPERGRAVITY }}
\footnote{ \em Presented at the conference on Gauge Theories,
Applied Supersymmetry and Quantum Gravity,
 Leuven, Belgium, July
1995. To be published in the conference proceedings}\\
\vspace{1.5cm}
Marcus T. Grisaru\footnote{
Work partially supported by the National Science Foundation under
grant PHY-92-22318.} \\
and\\
\vspace{1mm}
Marcia E. Wehlau\footnote{\hbox to \hsize{Current address:
Mars Scientific Consulting, 28 Limeridge Dr.,
 Kingston, ON CANADA K7K~6M3}}\\
\vspace{1mm}
{\em Physics Department, Brandeis University, Waltham, MA 02254, USA}

\vspace{1.1cm}
{{ABSTRACT}}
\end{center}

\begin{quote}
We describe various aspects of two-dimensional $N=2$ supergravity
 in superspace.
We present the solution to the constraints  in terms
of unconstrained prepotentials, and  the different superspace measures
(full and chiral) used in the construction of
invariant actions.  We discuss aspects of the theory in
light-cone gauge, including the Ward identities for correlation functions
defined with respect to the induced supergravity action.
\end{quote}

\vfill

\begin{flushleft}
September 1995

\end{flushleft}
\end{titlepage}

\newpage

\renewcommand{\thefootnote}{\arabic{footnote}}
\setcounter{footnote}{0}
\newpage
\pagenumbering{arabic}

\sect{Introduction}

Three years ago, P. van Nieuwenhuizen and one of the authors (M.T.G)
 performed some loop
calculations in Polyakov's quantum induced gravity in light-cone gauge
\cite{Poly}.
 We recall that away
from the critical dimension, the induced  gravity action takes the form
\beq
{\cal S}_{ind} =- \frac{ c}{24 \pi} \int d^2x R \frac{1}{\Box} R  \non
\eeq
which in light-cone gauge becomes
\bea
{\cal S}_{ind} &=& - \frac{ c}{24 \pi} \int
  d^2x \di_-^2 h_{++} \frac{1}{1- \frac{1}{\di_+} h_{++}
\di_-}  \frac{\di_-}{\di_+}
 h_{++}  \non\\
&=& - \frac{ c}{24 \pi} \int  d^2x \left[ h \frac{\di_-^3}{\di_+}h
 - h \left( \frac{\di_-^2}{\di_+} h \right)^2 + \cdots \right] \non ~~.
\eea

To determine the effective action, ${\cal S}_{eff}$, one computes
perturbatively
 in  powers of $1/c$.  The one-loop results are given in \cite{MP,MGPvN}.
In particular one finds renormalization $c \rightarrow Z_c c$
 and $h \rightarrow Z_h h$,   consistent with results in the
equivalent KPZ or WZNW formulation \cite{KPZ}, where one proves  that under
quantum
 corrections
\beq
\frac{c}{6} \rightarrow k+2 = \frac{1}{12} [c-1- \sqrt{(c-1)(c-25)}] \non
\eeq
As suggested by the equivalence, as well as other arguments
\cite{SSN,BG},  one expects a stronger result, namely the
 relationship between the effective and induced
actions is given by
\beq
{\cal S}_{eff} (h,c) = {\cal S}_{ind} (Z_h h, Z_c c) \non~~.
\eeq
 The work presented in \cite{MGPvN}
was done in part to understand perturbatively what happens to $c$ and to $h$
under renormalization (which  is understood differently in the
 KPZ or WZNW formulation),
and in part to gain some experience with quantum
calculations in non-local field theories.
The calculations turned out to exhibit some unpleasant features: first,
 there were
 ultraviolet linearly divergent loop integrals, so that results
 were in principle routing
dependent. Second,  and perhaps related to this first problem, it
 appeared that
the results were regularization dependent. The work used the
 Polyakov exponential
cutoff regularization, but calculations by other authors
  \cite{Sevrin}, using Pauli-Villars
regularization,  led to somewhat different results. At any rate,
 going beyond one loop
in order to check the relationship between $S_{eff}$ and $S_{ind}$ appeared
prohibitively difficult.

 It is conceivable  that the situation might be better in supergravity,
 where the ultraviolet behaviour is generally much improved, and
 supergraph techniques may simplify higher-loop
calculations. It turns out that  the ultraviolet behaviour is not sufficiently
improved
  in the case of  (1,0) or (1,1) supergravity \cite{MGRMX},
 but is for (2,0)  \cite{RMX}  or (2,2)
 supergravity. (However,
there are unpleasant features of (2,0) supergravity, such as anomalies
 induced by the lack
of left-right symmetry which make it a less appealing candidate for
 further investigation).
There are other aspects of Polyakov's induced gravity which one may
 wish to generalize to
the (2,2) case. Foremost are
the Ward identities which exhibit the hidden $SL(2, C)$ symmetry of induced
gravity and in principle
solve completely the theory.
These Ward identities  have also been used \cite{KKP} to  determine the
 dressing of  nonconformal matter  one-loop $\beta$-functions
 by induced gravity,
$\b_0 \rightarrow \b = \frac{k+2}{k+1} \b_0$, and a similar approach would
 be interesting  in the (2,2) supergravity case where perturbative methods
indicate the
 absence of such corrections \cite{MGDZ}.
Finally the (2,2) case is interesting in its own right because one believes
that Polyakov's light-cone results can be mapped onto the WZNW model, but for
 the (2,2) case
there are apparently problems constructing the WZNW action.

 In order to
study these questions, which are of a quantum nature,  one needs
 to have a formulation of $N=2$ supergravity in terms of unconstrained
 prepotentials which allows functional integration, etc..
Such a formulation has been absent in the past.  Recently,
  we have developed a prepotential approach to $N=2$ supergravity
which  allows discussion of the quantum properties of the theory,
 its light-cone formulation,
and the derivation of the Ward identities. In the following we
 summarize our procedure and
results. Further details can be found in the published references.

\sect{Solution of (2,2) Constraints}

  From the two-dimensional $N=2$ nonminimal
$U_A(1) \otimes U_V(1)$ supergravity theory,
 two versions of minimal (2,2) supergravity can be obtained \cite{HP,MGMW} --
 the axial
$U_A(1)$ version and the vector $U_V(1)$ version --
 depending upon which tangent space symmetry one
 gauges.  We focus on the axial version, which is related to
minimal four-dimensional $N=1$ supergravity by dimensional reduction.
 (Results
for the $U_V(1)$ case can be easily obtained from the axial solution.)

We work in a spinor light-cone basis
and use superspace coordinates
$ (x^{\pp} $, $x^{\mm}$ ; $\th^+$, $\th^-$, $\th^{\pd}$, $\th^{\md})$.  In the
nonminimal
theory the spinorial covariant derivatives are
\bea
\Del _{\a} &=& E_{\a}+ \Phi_{\a}{\cal M} + \S_{\a}' {\cal Y}' + \S_\a
{\cal Y}
\eea
where $\a=\pm$, with similar expressions for the complex conjugates,
as well as the  vectorial derivatives.
The tangent space Lorentz, $U_V(1)$ and $U_A(1)$
generators are ${\cal M}$, ${\cal Y}$ and ${\cal Y}'$, with the associated
connections
$\Phi_\a$, $\S_{\a}$ and $\S_{\a}'$.  The vielbein is given by
$E_A = {E_A}^M D_M$.

We list the constraints for the $U_A(1) \otimes U_V(1)$ theory \cite{HP,MGMW}.
They are
\bea
\{\Del_+ ,\Del_+\} &=&0 ~~~~~~~,~~~~  \{\Del_-, \Del_-\}=0 \nonumber\\
\{\Del_{+} , \Del_{\pd} \} &=&i \Del_{\pp} ~~~~,~~~~\{\Del_{-} , \Del_{\md} \}
= i \Del_{\mm}  ~~,
\eea
as well as
\bea
  \{\Del_+, \Del_-\} &=&
- \half \Rb ({\cal M} - i {\cal Y}')  \nonumber\\
\{\Del_+, \Del_{\md} \}&=& - \half \Fb ({\cal M} - i {\cal Y})
\ena
and their complex conjugates.

To find a description of one of the minimal $U(1)$ theories, we follow a
procedure similar to that used in the four dimensional case.  We start
with the {\em full} $ U_A(1) \otimes U_V(1)$ gauge group, and solve the
constraints for the nonminimal theory.
  The solution is derived in terms of two prepotentials, a real vector
superfield $H^m$ and a general complex superfield compensator $S$.
We obtain the solution in terms of the ``hat''
differential operators
\beq
\hat{E}_{\pm}= e^{-H}D_{\pm}e^{H}~~,~~ H=H^m i\pa_m  ~~,
\eeq
with the spinorial vielbein given by
\bea
E_+  &\equiv& e^{\Sb}(\Eh _+ + \Apm \Eh_-) \nonumber\\
E_- &\equiv&  e^{\Sb}(\Eh_-+\Amp \Eh_+) \label{Es}
\eea
(again with corresponding expressions for the complex conjugates).
 The quantities $A_{\a}^{~\b}$ in \reff{Es} can be solved for and
are given explicitly in terms of $H^m$ {\em only}, as is the vielbein
determinant $E$, whereas  the
 connections $\O_{\a}$, $\G_{\a}$ and $\S_{\a}$ are functions of
$H^m$, as well as $S$ and $\bar{S}$ \cite{MGMW}.

Now that we have these equations, we reduce the theory to one of its
minimal forms by ``degauging", i.e. by eliminating one of the (extra)
$U(1)$ symmetries.  To obtain the minimal $U_A(1)$ $(U_V(1))$ theory, we
restrict the gauge group by setting $\S_\a = 0$ ($\S_\a' = 0$), or
equivalently
the field strength $F = 0$ ($R = 0$).  Imposing this additional
restriction in the $U_A(1)$ case \cite{MGMW}, we find that the superfield
$\Sb$ can be expressed in terms of an arbitrary covariantly  antichiral
superfield $\bar{\s}$ as
\beq
e^{\Sb} = e^{\bar{\s}} \frac{ \left[1\cdot e^{-\BAH}
 \right]^{-\frac{1}{2}}}{[1-\Apm \Amp]^{\frac{1}{2}}} E^{-\frac{1}{2}}~~.
 \label{degauge}
\eeq
(Here $\BAH$ indicates that the
 differential operator in $H^m i \pa_m$ acts on objects to its left.)
This completes the degauging to the minimal $U_A(1)$ theory.

The unconstrained real vector superfield $H^m$ and the chiral scalar
superfield $\s$ are the prepotentials of minimal $U_A(1)$ (2,2)
supergravity.  To obtain the mirror image $U_V(1)$ theory, one simply
interchanges $-$ and $\md$ everywhere (as well as interchanging $R$
with $F$, and ${\cal Y}$ with ${\cal Y}'$).  This amounts to
replacing the chiral superfield $\s$ with a twisted chiral superfield
$\tilde{\s}$.

With this solution at hand we can now discuss invariant actions.
Details can be found in \cite{measures}, and
we simply list the results here.
Since
 the vielbein determinant depends solely on $H^m$, it is clear that we can
construct invariant actions in {\em full} superspace that are independent
of the compensator $\s$, such as
\beq
{\cal S} = \int d^2x d^4 \th E^{-1} {\cal L}
\eeq
where ${\cal L}$ is an arbitrary scalar function of superfields.

 For the minimal $U_A(1)$ theory we can rewrite the full superspace
integral above as as integral over chiral superspace,
\beq
{\cal S}= \int d^2x d^2 \th {\cal E}^{-1} \Delb^2 {\cal L}|_{\bar{\th}=0}
\label{chiral}
\eeq
with the chiral measure
\beq
{\cal E}^{-1} =  e^{-2\s} (1.e^{\BAH}) ~~.
\eeq

  The case for the $U_V(1)$ theory is similar. We can rewrite the full
superspace integral  as
\beq
{\cal S} = -\int d^2x d \th^+ d\th^{\md} \tilde{\cal E}^{-1} \Delpd \Delm
 {\cal L}|_{\tilde{\th}=0}
\eeq
with the twisted chiral measure
\beq
\tilde{\cal E}^{-1} =  e^{-2\tilde{\s}} (1.e^{\BAH}) ~~.
\eeq
These measures can be used for arbitrary (twisted) chiral integrands.

\sect{Transformation Laws}

 We now examine the gauge transformations of the prepotentials in  this
description of (2,2) supergravity.

In the solution of the constraints knowledge of the result in the
four-dimensional case
\cite{Superspace} was
used as a guide in eliminating many irrelevant superfields through
algebraic gauge transformations to specific supersymmetric gauges.
The desired form of $H= H^m i \di_m$
($H^m$
real and vectorial)
was determined by implicitly using the usual $K$ invariance of the
covariant derivatives ($\Del_A' = e^{iK} \Del_A e^{-iK}$, $K$ a real scalar
superfield), as well as some of the $\Lambda$ invariance which appears
as a result of the solution.  Specifically, it was possible to
gauge away the imaginary part of $H^m$, and  spinorial  superfields $H^\a$,
$H^{\dot{\a}}$. At this point some $\Lambda$ invariance still remains however,
and we
now discuss this, influenced again by experience with the four-dimensional
$N=1$ situation.

We start by examining the invariance under superspace coordinate
transformations of the kinetic action for covariantly chiral and antichiral
superfields \cite{lightcone}.
Covariantly chiral and antichiral scalar superfields $\Phi$, $\bar{\Phi}$
are defined by
$\Del_{\pmd} \Phi = \Del_{\pm} \bar{\Phi}=0$.  They can be
expressed in terms of ordinary chiral and antichiral superfields $\phi$,
$\bar{\phi}$, by
\beq
\Phi = e^H \phi e^{-H} ~~~,~~~ \bar{\Phi} = e^{-H}\bar{\phi}e^H ~~.
\eeq
Their  kinetic action  is
\bea
S&=& \int d^2x d^4 \th E^{-1} \bar{\Phi} \Phi \nonumber \\
&=&  \int d^2x d^4 \th E^{-1} \left(e^{-H} \bar{\phi}\right) e^{-H}\left(
e^{2H} \phi \right) \non \\
&=&\int d^2x d^4 \th E^{-1} e^{\BAH}\left( \bar{\phi} e^{2H} \phi \right)~~.
\nonumber
\eea
where we have performed some integration by parts.

Ordinary chiral and antichiral superfields transform under supercoordinate
transformations as
\beq
\phi \rightarrow  e^{i\L}\phi ~~~~,~~~ \bar{\phi} \rightarrow e^{i \bar{\L}}
\bar{\phi} \nonumber
\eeq
where $\L$ and $\bar{\L}$ are given by
\beq
\L=\L^m i\pa_m +\L^{\a}iD_{\a} +\L^{\ad}iD_{\ad}~~,~~
\bar{\L}=\bar{\L}^m i\pa_m +\bar{\L}^{\a}iD_{\a} +\bar{\L}^{\ad}iD_{\ad} ~~.
\eeq
The above action will be invariant under these transformations provided that
\beq
e^{2H} \rightarrow e^{i\bar{\L}} e^{2H} e^{-i\L}~~~,~~~
E^{-1} e^{\BAH} \rightarrow  E^{-1}e^{\BAH} e^{i \buildrel \leftarrow \over
\Lb} ~~.
\eeq
Similarly, by requiring invariance of the chiral integral in \reff{chiral} we
find
\beq
e^{-2\s} (1.e^{\BAH}) \rightarrow  e^{-2\s} (1.e^{\BAH}) e^{i \buildrel
\leftarrow \over
\L}
\eeq

The $\L$'s are restricted by two
requirements: they must
be (anti)chirality-preserving,  i.e. $D_{\pm} e^{i\Lmb} \bar{\phi}=0$
and they must maintain the vector nature of  the operator $H= H^m i\pa_m$.
These conditions imply that the $\L$'s can be expressed in terms of arbitrary
spinor parameters $L^\a$, $L^{\dot{\a}}$. The precise relation is given in
\cite{lightcone}.
In particular, at the linearized level,
 the transformations of the prepotentials are
\bea
\d H^{\pp}&=&\ihalf (D_{\md}L^+ -D_-{L}^{\pd} )  \nonumber\\
\d H^{\mm}&=&\ihalf (D_{\pd}L^--D_+{L}^{\md}) \nonumber\\
\d\s &=&-\ihalf \Db^2(D_+L^+-D_-L^- ) \nonumber\\
\d\bar{\s}&=&-\ihalf D^2(D_{\pd}{L}^{\pd}-D_{\md}{L}^{\md}) ~~.
\eea

\sect{Reaching Light-Cone Gauge}

Going to a specific gauge, where certain components of gauge fields ${\cal V}$
are
set to zero, involves examining their gauge transformations and showing that
for
any such transformation, $\cal{V} \rightarrow {\cal V}+\d {\cal V} = {\cal
V}+{\cal D}
{\cal L}$, one can solve for the gauge parameter ${\cal L}$ for any $\d {\cal
V}$.
We go to light-cone gauge by choosing $x^{\mm}$ as ``time'', so that
$1/\pa_{\pp}$ is
local and can be used when solving for gauge parameters without introducing
propagating
ghosts. We show that in $H^m$ it is possible to gauge away all of $H^{\mm}$ by
using the
gauge parameters $L^- $ and $L^{\md}$, and the
compensators $\s$, $\bar{\s}$, and certain components of $H^{\pp}$ by using
$L^+$ and $L^{\pd}$.  It is, of course,
sufficient to examine the linearized transformations.

We consider first the transformation
\beq
\d H^\mm = \ihalf (D_\pd L^--D_+L^\md ) ~~.
\eeq
 Component by component,
 $H^\mm$ can be gauged away completely by components of $L^-$ or
$L^\md$, with nothing left of $L^-$ to use in $\d \s \sim \bar{D}^2 D_- L^-$
to gauge away the compensator \cite{lightcone}.

We then look at the gauge transformations induced by $L^+$ and $L^\pd$ .
Using them to gauge away all of $H^\pp$ takes us to
superconformal gauge; instead we use them
 to gauge away the compensators, and some of the lower
components of $H^\pp$, by a Wess-Zumino gauge choice which eliminates
some left-over gauge invariance.
 We find thus a gauge in which
$\s =0$ and  the prepotential has the form
\bea
\lefteqn{H^{\pp} = \th^-\th^{\md}[h_{\mm}^{~\pp} +\th^+ \psi_{\md}^{~\pp}
-\th^{\pd}\psi_-^{~\pp}
-\th^+\th^{\pd}D^{\pp}]} \non \\
&&+\th^-[ e^{\sihalf \th^+\th^{\pd}\pa_{\pp}}(\l_-^{\pp}+\th^+N^{\pp})]
+\th^{\md}[e^{-\sihalf \th^+\th^{\pd}\pa_{\pp}}(\l_{\md}^{\pp}
+\th^{\pd}\bar{N}^{\pp})]
\eea
Therefore, in light-cone gauge $H^{\pp}$ has a decomposition
in terms of $(2,0)$ superfields, one of them real, the other two
chiral and antichiral, with respect to $D_\pd$ and $D_+$.
 Absorbing the explicit $\th^-$ and $\th^{\md}$ into the definition of these
superfields, we write
\beq
H^{\pp} \equiv {\cal H}^{\pp} +\chi^{\pp} +\bar{\chi}^{\pp}
 \non ~~,
\eeq
\beq
{\cal H}^{\pp}=\th^-\th^{\md} {\cal H}_{\mm}^{~\pp}~~,~~\chi^{\pp}=
\th^-\chi_-^{~\pp}
{}~~,~~ \bar{\chi}^{\pp}=\th^{\md} \chi_{\md}^{~\pp}~~.
\eeq

We now list, in light-cone gauge, the expressions for the various relevant
quantities in the theory.  We find it convenient to describe the results
in terms of the field
\beq
\calHtpp = \calHpp + \ihalf \xpp \dirl \xbpp ~~,
\eeq
and note that the dependence of the geometrical quantities on the
prepotentials is almost linear in this gauge:
\beq
E^{-1} = 1 - [D_\pd, D_+]\calHtpp -i \di_\pp(\x -\xb)^\pp
 +2 \pa_\pp \xpp \pa_\pp \xbpp +i D_+ \xpp \dirl D_\pd \xbpp
\eeq
\bea
e^{2\Sb}
&=&1 + 2D_+ D_\pd \calHtpp+ 2i \di_\pp \xbpp + \xpp {\di}_\pp^2 \xbpp
+ \di_\pp
      \xpp \di_\pp \xbpp + i D_+ \xpp {\dirl} D_\pd \xbpp  \\
e^{2S} &=& 1 - 2 D_\pd D_+ \calHtpp - 2i \di_\pp \xpp + \xbpp {\di}_\pp^2 \xpp
 + \di_\pp
    \xpp \di_\pp \xbpp + i D_+ \xpp {\dirl} D_\pd \xbpp
\eea
\bea
\Rb &=& -2 e^{-i(\x + \calHt) \di} D^2 [ 2i \di_\pp
 (\calHt + \x)^\pp - 4 \di_\pp\xpp\di_\pp\xbpp
        - \xpp {\di}_\pp^2 \xbpp - \xbpp {\di}_\pp^2 \xpp] \\
R &=& 2 e^{i(\xb + \calHt)\di} {\Db}^2 [-2i \di_\pp (\calHt + \xb)^\pp
 - 4 \di_\pp\xpp\di_\pp\xbpp
     - \xpp {\di}_\pp^2 \xbpp - \xbpp {\di}_\pp^2 \xpp] ~~.
\eea

\sect{Light-cone gauge transformations}

Ultimately we want to derive Ward identities for correlation functions
defined by (functional) averaging with the (nonlocal)
 induced supergravity action.  They are
 obtained from the invariance of the
functional integral under a change of variables which is a field
transformation. The
only requirement is that the variation of the
induced action be local, which can be achieved if the field
transformation is a
 gauge transformation for which the induced action is anomalous.
In our case this is true for the general gauge transformations of the
prepotentials, when restricted  to light-cone gauge and chosen to preserve the
 form of $H^{\pp}$.

We consider the general gauge transformation
\beq
e^{2H} \rightarrow e^{i \Lb} e^{2H} e^{-i\L} \nonumber
\eeq
choose  $L^- = L^{\md}=0$, and suitably restrict  $L^+$ and $L^{\pd}$.
This procedure is rather involved (details are given in
\cite{lightcone}) and we simply state the result here.
We find that the final form of the residual light-cone transformations is, with
arbitrary parameters $\a^{\pp}$, $\g_\md^{\pp}$ and $\bar{\g}_-^\pp$,
\bea
\d \xpp &=& iD_{\pd}[\g_\md^\pp  + \half \xpp D_+ \a^\pp
- \half  {\a}^\pp D_+ \xpp ] \nonumber\\
\d \xbpp &=& -i D_+ [ \bar{\g}_-^\pp +\half {\a}^\pp D_\pd \xbpp
-\half \xbpp D_\pd \a^\pp  ] \label{x}~~,
\eea
and also that a useful quantity which transforms simply is
\beq
 \check{\calH}^{\pp} = \calHpp +i \xpp \dirl \xbpp ~~,
\eeq
for which we have
\beq
\d \check{\calH}^{\pp} = -\half \th^- \th^\md \pa_\mm {\a}^\pp +
{\a}^\pp \pa_{\pp}\check{\calH}^{\pp}
 -iD_\pd {\a}^\pp
D_+ \check{\calH}^{\pp}
 -iD_+{\a}^\pp D_\pd \check{\calH}^\pp
 -\check{\calH}^{\pp}  \pa_{\pp}{\a}^\pp ~~. \label{Hc}
\eeq

\sect{Light-cone gauge Ward identities}

The induced $(2,2)$ supergravity action is given by
\beq
S_{ind}= \frac{c}{4\pi} \int d^6z \bar{R} \frac{1}{\Box_c}R
\eeq
where $\Box_c$ is the supergravity covariantized d'Alembertian. It is obtained
by coupling superconformal matter to supergravity. Integrating out the
matter gives rise, because of the superconformal anomaly, to this induced
action.

The correlation functions in the presence of the induced action are
defined by
\beq
<X(z_1,z_2,...z_n)>
= \int {\cal D} (H, \phi )e^{S_{ind}(H) +S_m(\phi ) }
X(z_1,z_2,...z_n) ~~,
\eeq
where
 $X(z_1,z_2,...z_n)$ denotes a  product of supergravity or matter
fields, and $S_m$ is the action for matter.
In the functional integral we make a change of
integration variables which is the residual $\L$-transformation defined in the
previous section, and
assume that $S_m$ is invariant, while the induced action varies into
the (local) anomaly.
We obtain the  Ward identity
\bea
0&=&  \int {\cal D} (H, \phi  )e^{S_{ind}(H) +S_m(\phi)} \left[ \d S_{ind}
       X(z_1,z_2,...z_n)
      + \sum _i \d _i X(z_1,z_2,...z_n) \right] \nonumber\\
&=& < \d S_{ind} X(z_1,z_2,...z_n) > + \sum_i
    < \d_i X(z_1,z_2,...z_n)> ~~,
\eea
where $\d_i X(z_1,z_2,...z_n)$ is the variation of the $i$'th field in
$X(z_1,z_2,...z_n)$.
The functional integration in light-cone gauge is over
 $\xpp$, $\xbpp$ and $\calHpp$, with the gauge transformations
given in \reff{x} and \reff{Hc}.
For the matter fields, e.g.  chiral scalar
superfields with weight $\l$, we assume transformations such as
$\d \phi = i[\L, \phi ] +i \l(1.{\buildrel \leftarrow \over \L} )\phi$,
i.e. $\d \phi =  -i\bar{D}^2 (L^+D_+ \phi )+i \l (\bar{D}^2 D_+ L^+)\phi$.

The variation of the induced action is
\bea
\d S_{ind}&=& \frac{ic}{\pi} \int d^2x d^4 \th
 \left\{\left(i[D_+,D_\pd] \di_\pp \check{\calH}^\pp
 +\di_\pp^2 (\xpp -\xbpp) \right)\a^\pp \right.   \non \\
&& {~~~~~~} \left. +2 \pa_\pp^2 \xpp D_+\bar{\g}_-^\pp
 - 2\pa_\pp^2 \xbpp D_\pd \g_\md^\pp \right\} ~~,
\eea
and we note that  $\a$ and $\g$ are arbitrary gauge parameters.

We  can verify, as an explicit example in perturbation
theory,  the Ward identity for the correlator $<\phib (y) \phi(z)>$ for an
ordinary
chiral scalar superfield ($\l =0$). The matter action is
\bea
S_m  &=&\int d^6z E^{-1}
\left(e^{-H}\bar{\phi}\right) \left(  e^H \phi \right) \nonumber\\
&=& \int d^6z
( \bar{\phi} \phi -2 H^\pp D_\pd \bar{\phi} D_+ \phi +\cdots ) ~~.
\eea
The  Ward identity is
\beq
<\d S_{ind} \fb \f> + <\d \fb \phi> + < \fb \d \phi > = 0
\eeq
with
\bea
\d \phi &=& -i{\Db}^2 [\th^\md (\a^\pp - 2i \xpp \di_\pp \a^\pp
 +D_\pd D_+  (\xpp \a^\pp )+ 2 D_\pd
           \g_\md^\pp) D_+ \phi] \non\\
\d \fb &=& -i D^2 [\th^-(\a^\pp + 2i \xbpp \di_\pp \a^\pp
 -D_+D_\pd  (\xbpp \a^\pp )+ 2 D_+ \gb_-^\pp) D_\pd \fb] ~~.
\eea
We note that terms linear in $c$ trivially satisfy the Ward identity,
terms independent of $c$ lead to tree graphs and
 terms proportional to $1/c^L$ give rise to loops.  Obviously the
terms that depend on $\a$ and $\g$ must separately satisfy the Ward
identity.
We consider  the $\a$-dependent part and obtain
\bea
\lefteqn{\frac{ic}{\pi} \int d^2x d^4 \th <i[D_+, D_\pd] \di_\pp \calH^\pp(x)
 \a^\pp(x) \fb(y) \f(z)>}  \non \\
&&-i <D^2 [\th^- \a^\pp D_\pd \fb(y)] \f(z)> -i <\fb(y) \Db^2 [\th^\md \a^\pp
D_+ \f(z)]> = 0
\eea
which we  have verified by an explicit  tree-level calculation in
\cite{lightcone}.

\sect{Comments}

We have outlined the description of (2,2) supergravity in terms of
unconstrained prepotentials, and  its light-cone properties.
We note that the choice of a {\em ghost-free} light-cone gauge, with no
residual gauge invariance for the induced action,
necessarily forces us to a Wess-Zumino-type gauge, which does not
have manifest (2,2) supersymmetry. In this respect the situation
is different from the case of the other supergravity theories that
have been studied. However, the lack of manifest (2,2) supersymmetry
is not a significant impediment to most applications.

The primary application of the light-cone formulation of (2,2) supergravity
is the investigation of  properties of the induced non-local action
$S_{ind}$.  Specifically, one can generalize to the (2,2) case
Polyakov's $SL(2,C)$ symmetry for induced gravity, as was done for
(1,0) and (1,1) supergravity in \cite{MGRMX} and for  (2,0) supergravity in
\cite{RMX}.   Comparison of our results with those of this last
reference, and in particular the transformation    in (5.4),
shows that our Ward identities  involving $ \calH^\pp$  are the same as
those in the (2,0) case
so that, upon  solving them, one expects to find the same
 associated  $OSp(2|2)$ current algebra.  However, in  the (2,2) theory
which involves the superfields $\chi$, $\bar{\chi}$ in addition to  ${\cal H}$,
one avoids the anomalies due to the inherent left-right asymmetry
of the (2,0) theory.

The bosonic Ward identities have been used to determine the induced gravity
corrections to the $\beta$-functions in non-conformal theories
 \cite{AJ,KKP,Sch}. One
of the authors (M.T.G.), with D.~Zanon \cite{MGDZ}, has examined
supergravitational dressing in perturbation theory.  Corrections to the
$N=1$ $\s$-model $\b$-functions were found, but there were no one-loop
corrections in
the $N=2$ case.  It is of interest  to recover these results by using the Ward
identities
we have derived.

 Lastly, one can attempt to do higher loop calculations in the induced
supergravity
in light-cone gauge.  At one-loop, one gets a result consistent with
$c-1 = 2(k+1)$ \cite{PZ}.  As mentioned in the introduction, higher loop
calculations
generalizing those of \cite{MGPvN} may be feasible, and free of the problems
encountered in the bosonic case.


\begin{thebibliography}{99}
\bibitem{Poly} A.M. Polyakov, Int. J. Mod. Phys. {\bf A2} (1987) 893.
\bibitem{MP} K.A. Meissner and J. Pawelczyk, Mod. Phys. Lett. {\bf
A5} (1990) 763.
\bibitem{MGPvN} M.T. Grisaru and P.~van Niewenhuizen, Int. J. Mod. Phys.
{\bf A7} (1992) 5891.
\bibitem{KPZ}  V. Knizhnik, A.M. Polyakov and A.B. Zamolodchikov,
Mod. Phys. Lett. {\bf A3} (1088) 819.
\bibitem{SSN} K. Schoutens, A. Sevrin and P. van Nieuwenhuizen, Nucl. Phys.
{\bf B364} (1991) 584.
\bibitem{BG} J. de Boer and J. Goeree, Nucl. Phys. {\bf B401} (1993) 348.
\bibitem{Sevrin} A. Sevrin, R. Siebelink and W. Troost, Nucl. Phys. {\bf B413}
(1994) 185.
\bibitem{MGRMX} M.T. Grisaru and R.M. Xu,  Phys. Lett. {\bf B205} (1988) 486;
N.D. Hari Dass and R. Sumitra, Int. J. Mod. Phys. {\bf A4} (1989) 2245.
\bibitem{RMX} R.-M. Xu, Phys. Lett. {\bf 247} (1990) 295; W. Sabra,
Int. J. Mod. Phys. {\bf A6} (1991) 755.
\bibitem{KKP} I.R. Klebanov, I.I. Kogan and A.M. Polyakov, Phys. Rev. Lett.
{\bf 71} (1993) 3243.
\bibitem{MGDZ} M.T. Grisaru and D. Zanon,  Phys. Lett. {\bf B353} (1995) 64.
\bibitem{HP}  P. Howe and  G. Papadopoulos, Class. Quantum Gravity
{\bf 4} (1987) 11; S. J. Gates, L. Liu and R. Oerter, Phys. Lett. {\bf B218}
(1989) 33.
\bibitem{MGMW} M.T. Grisaru and M.E. Wehlau, Int. J. Mod. Phys. {\bf A10}
(1995) 753.
\bibitem{measures} M.T. Grisaru and M.E. Wehlau, {\em Superspace Measures,
 Invariant Actions, and Component Projection Formulae for (2,2)
Supergravity}, hep-th/9508139.
\bibitem{Superspace} S.J. Gates, Jr., M.T. Grisaru, M. Ro\v{c}ek  and W.
Siegel,
{\em Superspace} (Benjamin-Cummings, 1983).
\bibitem{lightcone} M.T. Grisaru and M.E. Wehlau, {\em (2,2) Supergravity
in the Light-Cone Gauge}, hep-th/9505068, Nucl. Phys. B, to be published.
\bibitem{AJ}  J. Ambjorn and K. Ghoroku, Int. J. Mod. Phys. {\bf A9} (1994)
5689.
\bibitem{Sch}  C. Schmidhuber, Nucl. Phys. {\bf B404} (1993) 342.
\bibitem{PZ} A.M. Polyakov and A.B. Zamolodchikov, Mod. Phys. Lett. {\bf A3}
(1988) 1213.
\end{thebibliography}
\end{document}